\documentclass[12pt,preprint]{aastex}
\shortauthors{Raichur and Paul}
\shorttitle{QPOs in Cen X-3}

\slugcomment{To be submitted to The Astrophysical Journal}
\received{}

\begin{document}

\title{Quasiperiodic oscillations in Cen X-3 and the long term
intensity variations}

\author{Harsha Raichur\altaffilmark{1,2} and Biswajit Paul\altaffilmark{1}}

\altaffiltext{1}{Raman Research Institute, Bangalore 560\,080, India}
\altaffiltext{2}{Joint Astronomy Programme, Indian Institute of Science, Bangalore 560\,080, India}

\begin{abstract}

We have investigated properties of the Quasi Periodic Oscillation (QPO)
features in the accretion powered X-ray pulsar Cen X-3 over a period of
about four years using observations carried out with the Proportional Counter 
Array (PCA) of the {\it {Rossi X-ray Timing Explorer}}. The observations cover
a wide range of X-ray intensity of the source in excess of the binary
intensity modulation. We have detected QPOs in
11 out of a total 81 pointings with the PCA
with rms intensity fluctuation upto 10\%. The QPO peak
frequency shows clustering around 40 and 90 mHz with the QPO
frequency having no dependence on X-ray intensity.
This indicates that either (a) the observed X-ray
luminosity of the source is not related to the mass accretion rate or
inner radius of the accretion disk or (b) that the QPO generation
mechanism in Cen X-3 is different from the beat frequency model or
Keplerian frequency model that is believed to be operational in most
other transient and persistent X-ray pulsars. We have also found that,
the rms variation in the 40 mHz QPO feature is not dependent on the
X-ray energy, indicating that disk absorption related origin for the
QPO is unlikely.  

\end{abstract}

\section{Introduction}

The lightcurves of X-ray binary pulsars show periodic intensity variations
with the spin of the neutron star and its orbital motion. But a few of the
persistent X-ray binary pulsars also show a long term periodic intensity
variation with time scales more than an order of magnitude greater than the
orbital period of the binary. 
Periodic superorbital intensity variations are seen in Her X-1 (35 day: Still 
\& Boyd 2004), LMC X-4 (30.5 day: Paul \& Kitamoto 2002) and 2S 0114+650 (30.7 
day: Farrell et al 2006). SMC X-1 shows quasi periodic superorbital intensity 
variations with a 50-60 day cycle (Clarkson et al. 2003).   
The intensity variations in 
these systems are understood to be due to obscuration of the central X-ray 
source by a warped precessing accretion disc. Spectral studies of Her X-1 and
LMC X-4 show iron line intensity and equivalent width evolving during the 
superorbital periods. Also the absorption column density 
in the line of sight is found to be higher during the low intensity states 
indicating an excess of absorbing matter in the line of sight during these 
times (Naik \& Paul 2003).

Cen X-3 is a high mass X-ray binary pulsar with very strong but aperiodic
long term intensity variations (Figure 1). This is the first X-ray pulsar 
discovered (Giaconni et al. 1971) and is also the brightest persistent pulsar. 
It has a spin period of $\sim$4.8 s and an overall spin-up trend with alternate
spin-up and spin-down intervals which last from about 10 to 100 days
(Finger, Wilson \& Fishman 1994). It has an orbital period of 2.1 days and a 
companion star
of about $20M_{\odot}$ (Avni \& Bahcall 1974). Though Cen X-3 is a persistent 
pulsar, its binary period averaged X-ray intensity varies by a factor of more 
than 40 (Paul, Raichur \& Mukherjee 2005).
As the long term intensity variation of 
Cen X-3 does not remotely appear to have any periodic or quasi-periodic nature 
(Paul et al. 2005), it is natural to assume that the X-ray flux variation
is due to changing mass accretion rate.
However, using a strong dependence of the orbital modulation and the pulsed
fraction of Cen X-3 on its X-ray intensity state we have shown that
the long term X-ray intensity variation in this source can be
due to change in obscuration by an aperiodically precessing warped accretion
disk (Raichur and Paul 2008). In this scenario, as the X-ray intensity
decreases, reprocessed and unpulsed X-rays from a relatively large scattering
medium progressively dominates the observed X-ray intensity.
We further investigate this hypothesis using the Quasi Periodic Oscillations
(QPO) in Cen X-3 with respect to its intensity state. In the accretion
powered X-ray pulsars, the QPOs are
understood to be due to inhomogenities in the inner accretion disk and
therefore its frequency is expected to be related to the inner radius of the
accretion disk. A correlation between the QPO frequency and X-ray luminosity
(and hence mass accretion rate / inner disk radius) has been observed in
several transient and persistent X-ray sources which show a large range of
X-ray intensity (EXO 2030+375: Angelini, Stella \& Parmar 1989, 3A 0535+262:
Finger, Wilson \& Harmon 1996, XTE J1858+034: Mukherjee et al. 2006, 4U 
1626--67: Kaur et al. 2008).  

QPOs are known to be present in all types of accreting X-ray pulsars.
In most sources it is a transient phenonema and QPOs have been detected
in about a dozen out of about a hundred known accreting X-ray pulsars.
With a few exceptions (4U 1748-288: Zhang et al. 1996 and XTE J 0111.2-7317:
Kaur et al. 2007) the QPO frequency is ususally in the range of 40-200 mHz,
consistent with it being related to the inner radius of the accretion disk
around a highly magnetised neutron star in its bright X-ray state.
Previous studies of Cen X-3 power spectrum have shown Quasi Periodic 
oscillation (QPO) at $\sim 40$ mHz (Takeshima et al. 1991).

In the present work, we have mainly studied the QPOs of Cen X-3 and its
relation to the source intensity if any. For this we have analysed all the
available archival data of the {\it{Rossi X-ray Timing Explorer (RXTE)}}
proportional courter array (PCA) from the year 1996 to 2000.    
   
\section{Observations and data analysis}

Cen X-3 was observed extensively by {\it RXTE}-PCA during 1996-1998 and again
for some time in 2000. We have analysed X-ray lightcurves from all the data
available during this period. A total of 525 ks data was obtained from 81
pointings carried out in
this period. Very few of the observations were carried out during the eclipse
or ingres/egress of eclipse and data collected during these periods were
excluded from further analysis. Table 1 gives details of the observations.

Light curves were extracted from all the observations using Standard-I data
which has a time resolution of 0.125 s. Power spectrum was obtained for 
each of these observations using the Standard-I lightcurve from data
streches of duration 1024 s. Power spectra from all such stretches within
one observation pointing were averaged and normalised such 
that their integral gives the squared rms fractional variability and the 
expected white noise level was subtracted. We have detected QPO features
at different frequencies as described below. However, QPO features were 
not present in all the data sets. Table 2 lists mid-times of the 11 segments
in which QPOs were detected along with the QPO frequencies. We have detected
the earlier reported QPOs around 40 mHz in all of the years although not in
every observation. The QPO at 90 mHz was seen only in the observations made
in 1996 and never again. The two QPOs 40 mHz and 90 mHz are not seen
together. In all the power spectra in which the QPOs were detected, the
peak associated with pulsar spin period was seen very clearly at $\sim 0.2$
Hz along with several harmonics. As reported earlier, pulsations were not
detected below a background subtracted count rate of 50 per proportional
counter unit with an upper limit of 0.8 \% on the pulsed fraction (Raichur
and Paul 2008). Some representative power spectra in different intensity
states, with and without the QPO features are shown in Figure 2 and
Figure 3.

Another feature seen in most of the power spectra is a broadening of the
fundamental spin frequency peak. To investigate the reason for this broadening
we chose the 
observation which had the highest RMS of the spin frequency broadening feature 
(Obs. Id P20104, TJD of observation 10508.012 - 10508.713). The total length of
this observation is about 60608 s and it gives 51 intervals of 
1024 s each. A final power spectrum using 1024 s of data streches averaged over 
all the intervals was made and then fitted with the model power spectrum given 
by Lazzati and Stella (1997). The model power spectrum considers the fact that
any aperiodic variability in the emission from accretion column(s) 
of a magnetic neutron star should be modulated at the X-ray pulsar period,
hence giving rise to a coupling between the periodic and aperiodic variability
(Burderi et al. 1997; Menna et al. 2003).
The coupling parameter R as defined in the model, is a measure of the degree 
of coupling between the periodic and the aperiodic variabilities. The effects
due to a finite length of the lightcurve are built into the model. Fitting
this model (equation 5 of Lazzati and Stella 1997)
to our power spectrum gives $R \approx 0.64$, similar to the results
derived by Lazzati and Stella (1997) using EXOSAT data for Cen X-3. A value of
R $\approx$ 1.0 or greater indicates a strong coupling between the aperiodic 
and periodic variabilities.

In the top panel of Figure 4, a plot of the QPO frequency is shown against
the 2-30 keV X-ray flux measured with the PCA. To determine the X-ray fluxes
we have fitted the X-ray spectra with a simple model consisting of a high
energy cut-off power law along with line of sight absorption and iron emission
lines. Part of the X-ray light curves that showed pre-eclipse X-ray dips were
excluded from X-ray flux determination.
The PCA X-ray light curves from which the power spectra have been
generated represent the instantaneous measurement of X-ray flux over a
small period of a few ks. However, even outside the X-ray eclipse, within one
orbital period, the X-ray intensity of Cen X-3 varies smoothly by more than a
factor of two and rapidly by a factor of upto 4 during the pre-eclipse dips.
In a circular orbit, the orbital intensity variation is due to
different visibility of the X-ray source and its reprocessing region rather
than changing mass accretion rate. Therefore, we have also looked at the
QPO frequency against the orbital phase averaged X-ray intensity using data
from the All Sky Monitor (ASM) onboard {\it RXTE}. ASM has three detectors
which scan the sky
in a series of 90 s dwells in 3 energy bands, namely 1.5-3, 3-5 and 5-12 keV
(Levine, et. al.,1996). The combined ASM lightcurve with about 10-20 exposures
during each binary orbit of Cen X-3 gives a better estimate of the overall
X-ray intensity state of the source. We used the corresponding binary period
averaged count rates from ASM lightcurve to see any dependence of the
QPO frequency on the X-ray intensity.
The ASM count rates given in Table 2 are obtained from orbital period
averaged lightcurve after removal of data taken during the eclipse.
A plot of the QPO frequency with ASM count rate is shown in the bottom
panel of Figure 4. It is very clear from the figure that there is no
dependence of QPO frequency with the instantaneous or orbital phase averaged
X-ray intensity. The QPO features are clustered around two frequencies,
40 mHz and 90 mHz. Even within each cluster, there is no dependence of
QPO frequency with the X-ray intensity.
In the top panel of Figure 4 we have also shown a plot of expected QPO
frequency as a function of the X-ray flux in the  beat frequency model.
To calculate the X-ray flux expected by the beat frequency model we have taken 
a source distance of 8 kpc (Krzeminski 1974), and a magnetic field strength of 
$3.4 \times 10^{12}$ G (Coburn et. al. 2002).

We have also carried out an energy resolved QPO analysis from one of the
observations (ID P10132) in which strong QPOs were detected at $\sim 40$ mHz
using energy resolved binned mode and event mode data. We extracted 
lightcurves with time resolution of 0.125 s in energy bands of 2-4.1, 
4.1-6.6, 6.6-9.5, 9.5-13.1, 13.1-16.7, 16.7-20.4, 20.4-25.7 and
25.7-34.8 keV. The energy resolved analysis of the 40 mHz QPO, as
shown in Figure 5 did not reveal
any measurable dependence of the rms fractional variability on energy.
Energy resolved analysis for the 90 mHz QPO could not
be done as the data from the corresponding observation did not have the 
required energy and timing resolution in any of the data storage modes.

\section{Results and discussions}

We can summarize the principal results of our analysis presented in the 
previous section as follows:

1. Cen X-3 shows intermittent QPOs in different frequency ranges namely 40 mHz, and 90 mHz.

2. The presence of the QPOs or the frequency of the QPOs are not  
related to the luminosity state of the source. The RMS fluctuations associated 
with the QPOs are not correlated with the luminosity of the source. 

3. RMS fluctuation of the 40 mHz QPO is energy independent.

4. A weak coupling is measured between the low frequency aperiodic
variabilities and the spin frequency.

In the discussion that follows we argue that the observed QPO properties
of Cen X-3 is in agreement with the scenario in which the long term X-ray
intensity variation is due to change in obscuration by an aperiodically
precessing warped accretion disk (Raichur and Paul 2008).

The radius of the inner accretion disk around a magnetised neutron star with a 
mass of 1.4 $M_{\odot}$ and a radius of 10 km can also be approximately 
expressed in terms of its magnetic moment and X-ray luminosity as
(Frank, King and Raine  1992)
\begin{equation}
r_M = 3 \times 10^8 L_{37}^{-2/7}\mu_{30}^{4/7}
\end{equation}
where $L_{37}$ 
is the X-ray luminosity in the units of $10^{37}$ erg and $\mu_{30}$ is the 
magnetic moment in units of $10^{30}$ G cm$^3 $. For disc accretion,
often a scaling factor of 0.5 is used with the above expression of $r_M$. 
Then the radius of the inner accretion disk would be $R_M = 0.5 r_M$.  
Coburn et. al. (2002) have 
estimated the magnetic field of Cen X-3 neutron star to be 
$B \simeq 3.4 \times 10^{12}$ G (i.e, $\mu_{30} =  3.4$) using the cyclotron
absorption line in the X-ray spectrum of the source. 

The lowest and the highest 3-30 keV X-ray flux at which the 40 mHz QPO feature
is seen are $1.1 \times 10^{-9}$ and $1.18 \times 10^{-8}$ erg 
cm$^{-2}$s$^{-1}$ respectively. Assuming a distance of 8 kpc (Krzeminski 1974),
these correspond to X-ray luminosity of $L_{low} = 2.64 \times 10^{37}$ erg 
s$^{-1}$ and $L_{high} = 2.83 \times 10^{38}$ erg s$^{-1}$. If the observed 
X-ray luminosity represents the true X-ray luminosity and a proportional
mass accretion rate of Cen X-3, the inner accretion disk radius ($R_M$) will
approximately vary between $3 \times 10^8$ cm and $1.5 \times 10^8$ cm.
We note that the corotation radius of Cen X-3 ($P_{spin} \sim 4.8$ s) is
$4.7 \times 10^8 cm$, larger than the inner disk radius for the lowest X-ray
luminosity, and the QPO detections are outside a possible propeller regime.

The two largely used QPO models are the Beat Frequency model (BFM) and the 
Keplarian Frequency model (KFM). The BFM explains the QPO as the beat between 
the spin frequency $\nu_{spin}$ and the keplarian frequency $\nu_k$ of the 
inner accretion disk $\nu_{QPO} = \nu_{k}-\nu_{spin}$. Thus the radius of 
the inner accretion disk according the BFM is given as follows.
\begin{equation}
r_{M,BFM}=\left(\frac{GM_{NS}}{4\pi^2(\nu_{spin}+\nu_{QPO})^2}\right)^{1/3}
\end{equation}
In KFM, the QPO occurs at the keplarian frequency of the inner accretion 
disk $\nu_{QPO} = \nu_{k}$. Then radius of the inner accretion disk due to KFM 
will be
\begin{equation}
r_{M,KFM}=\left(\frac{GM_{NS}}{4\pi^2\nu_{QPO}^2}\right)^{1/3}
\end{equation}

However, in the
case of Cen X-3, as the $\nu_{spin}$ is larger than the observed QPO
frquencies, KFM is not applicable. This is because if the inner accretion disk
rotates slower than the neutron star, propeller effect is expected to inhibit
accretion of material from the accretion disk. Thus assuming an inner accretion
disk origin of the QPOs and using equations 1 and 2, one can experss a
relation between the QPO frequency and the X-ray luminosity. In the top
panel of Figure 4, we have shown the expected QPO frequency against the
measured X-ray for a source distance of 8 kpc. It is obvious from the figure
that the QPO frequency of Cen X-3 does not have the flux dependence as
expected in the beat frequency model.

From a study of the X-ray intensity dependence of the orbital modulation and
pulsed fraction in Cen X-3, recently we have proposed that the different flux
states of Cen X-3 are primarily due to varying degree of obscuration by an
aperiodically precessing warped accretion disk (Raichur and Paul 2008). The 
nearly constant QPO frequency (ignoring the rare 90 mHz feature) reported here
is indeed consistent with this hypothesis. We propose that the mass accretion
rate and thus the inner accretion disk radius of Cen X-3 is not highly variable,
thus the source produces a nearly constant QPO frequency. We note here that
the frequencies predicted by the BFM are significantly larger than the measured
ones. However, the expression used here for magnetospheric radius is only approximate
and a different prescription for the magnetospheric radius (for example if it is
considerably larger than $r_M$ given in equation 1) can explain the observed QPO
frequencies at the highest observed X-ray flux state and also be consistent with
the proposal that the QPO frequency is insensitive to the measured X-ray flux as
the X-ray flux variation is primarily due to disk obscuration.

We also note the other possibility that the QPOs in Cen X-3 may
not be due to any material inhomogenity in the inner accretion disk as is the
case for a few other X-ray binary pulsars like A0535+262 (Finger et al.
1996), EXO 2030+375 (Angelini et al. 1989) XTE J1858+034 (Mukherjee et al.
2006) and 4U 1626--67 (Kaur et al. 2008). In these transient binary X-ray
pulsars, the QPO frequency is well or
somewhat correlated with the X-ray luminosity of the source and hence the QPO
frequency variations are understood to be due to changes in the mass accretion
rate and associated changes in the radius of the inner accretion disk.

One study which could
give us more insight would be to study the emission lines from neutral, H-like 
and He-like iron atoms in the photoionised circumstellar material. Such  
spectral observations done during the eclipse ingress and egress of the source 
would give us knowledge of the distance at which the lines are produced. 
Measurements carried out at different intensity states of the source will tell 
us if the observed X-ray intensity is a true measure of the X-ray luminosity 
and the ionization parameter.

\acknowledgements
We thank an anonymous referee for many suggestions that helped us to improve
the paper.
This research has made use of data obtained from the High Energy Astrophysics 
Science Archive Research Center (HEASARC), provided by NASA's Goddard Space 
Flight Centre.

{}

\clearpage

{
\begin{table}
\caption{List of Observations}
~\\
\begin{tabular}{|c|c|c|c|}
\hline
Year&Obs Ids&No. of Pointings&Total Durations(ks)\\
\hline
1996&P10063&3&40\\
    &P10132&5&18\\
    &P10133&4&92\\
    &P10134&7&143\\
    &P10144&5&32\\
\hline
1997&P20104&12&32\\
    &P20105&3&26\\
    &P20106&7&14\\
\hline
1998&P30083&10&13.6\\
    &P30084&23&65\\
\hline
2000&P40072&2&50\\
\hline
\end{tabular}
\end{table}
}

\begin{table}
{
\caption{Details of QPOs detected in Cen X-3}
~\\
\begin{tabular}{|c|c|c|c|c|}
\hline
Mid Time of&ASM Count&QPO Freq&QPO&RMS\\
Observation&Rate&(Hz)&width&\\
(MJD)&&&&\\
\hline
50146.7283&$17.52\pm1.21$&$0.0903\pm0.0013$&$0.0091\pm0.0012$&$10.95\pm1.93$\\
50147.3781&$17.52\pm1.21$&$0.0931\pm0.0015$&$0.0094\pm0.0016$&$ 6.21\pm1.52$\\
50345.6888&$17.12\pm1.35$&$0.0452\pm0.0010$&$0.0068\pm0.0010$&$10.88\pm2.23$\\
50509.5281&$13.60\pm0.92$&$0.0384\pm0.0008$&$0.0113\pm0.0009$&$5.11\pm0.46$\\
50991.4068&$10.09\pm1.80$&$0.0454\pm0.0019$&$0.0125\pm0.0027$&$ 5.71\pm1.19$\\
50998.7648&$17.73\pm1.47$&$0.0403\pm0.0016$&$0.1281\pm0.0026$&$ 6.19\pm1.03$\\
50999.7818&$17.73\pm1.47$&$0.0419\pm0.0019$&$0.0048\pm0.0036$&$ 3.43\pm1.18$\\
51094.9594&$ 1.83\pm1.21$&$0.0431\pm0.0009$&$0.0069\pm0.0010$&$ 5.93\pm1.16$\\
51095.9637&$ 1.83\pm1.21$&$0.0445\pm0.0015$&$0.0064\pm0.0015$&$ 4.93\pm1.57$\\
51578.9780&$ 8.97\pm1.19$&$0.0402\pm0.0016$&$0.0051\pm0.0013$&$ 3.75\pm1.01$\\
51579.9307&$ 7.06\pm1.06$&$0.0502\pm0.0022$&$0.0101\pm0.0018$&$ 6.49\pm1.29$\\
\hline
\end{tabular}
}
\end{table}

\clearpage

\begin{figure}
\centering
\includegraphics[height=5in, angle=-90]{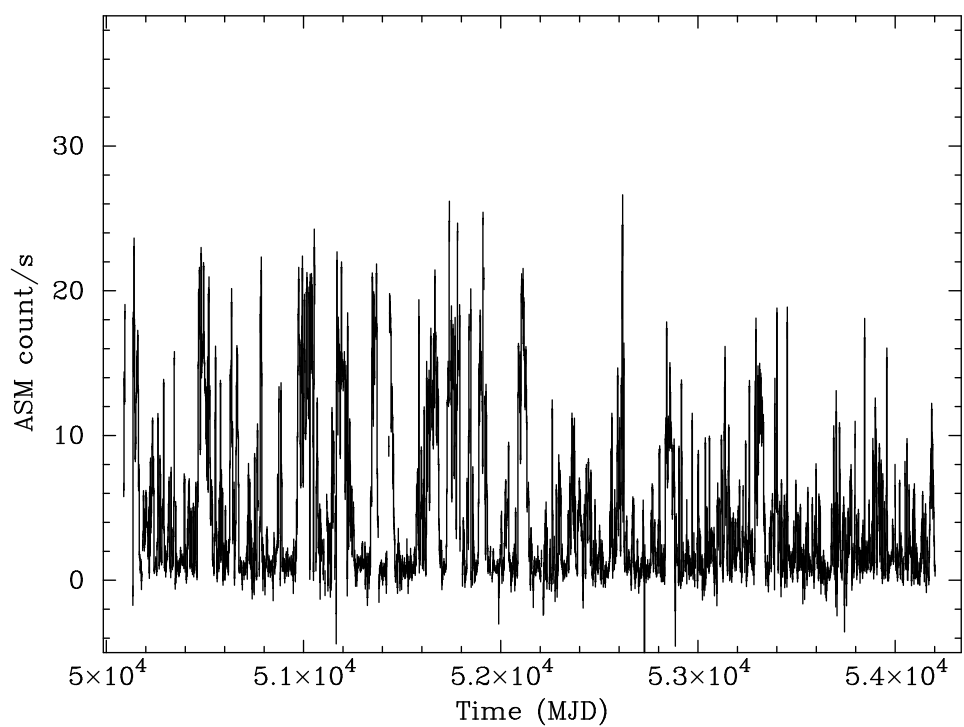}
\caption{The 1.5-12.0 keV band {\it RXTE}-ASM lightcurve of Cen X-3 is shown
here with a binsize same as the orbital period.}
\end{figure}

\begin{figure}
\centering
\includegraphics[height=5in, angle=-90]{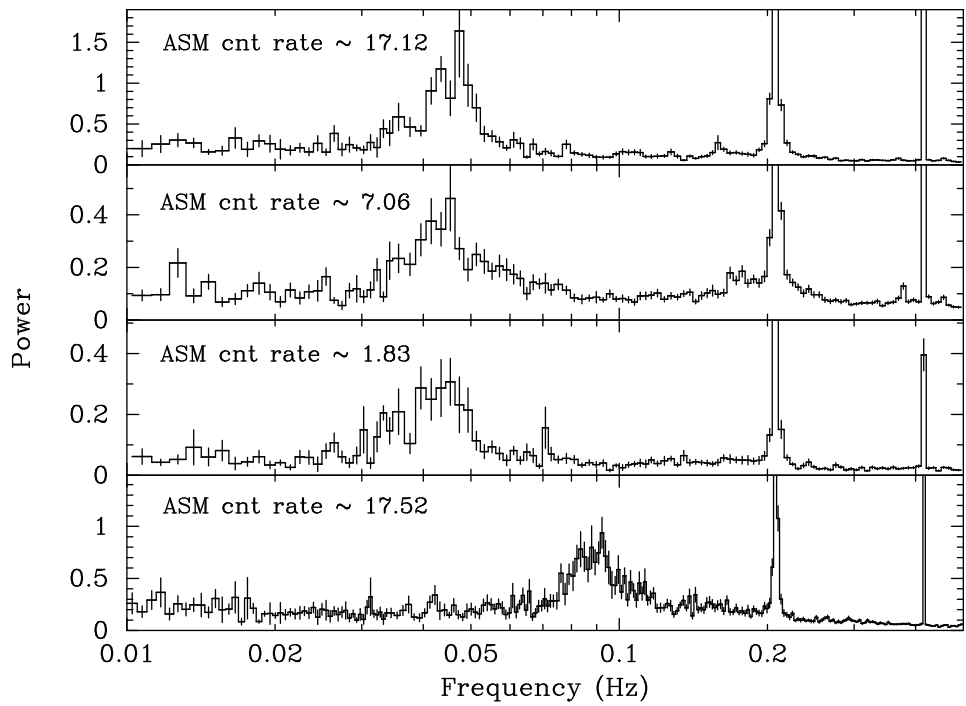}
\caption{Representative power spectra of en X-3 with QPOs are shwon here. 
Top three pannels show the 40 mHz QPO at different intensity states of Cen X-3.
The bottom pannel shows the 90 mHz QPO seen only during the high intensity 
state of Cen X-3 in 1996.}
\end{figure}

\begin{figure}
\centering
\includegraphics[height=5in, angle=-90]{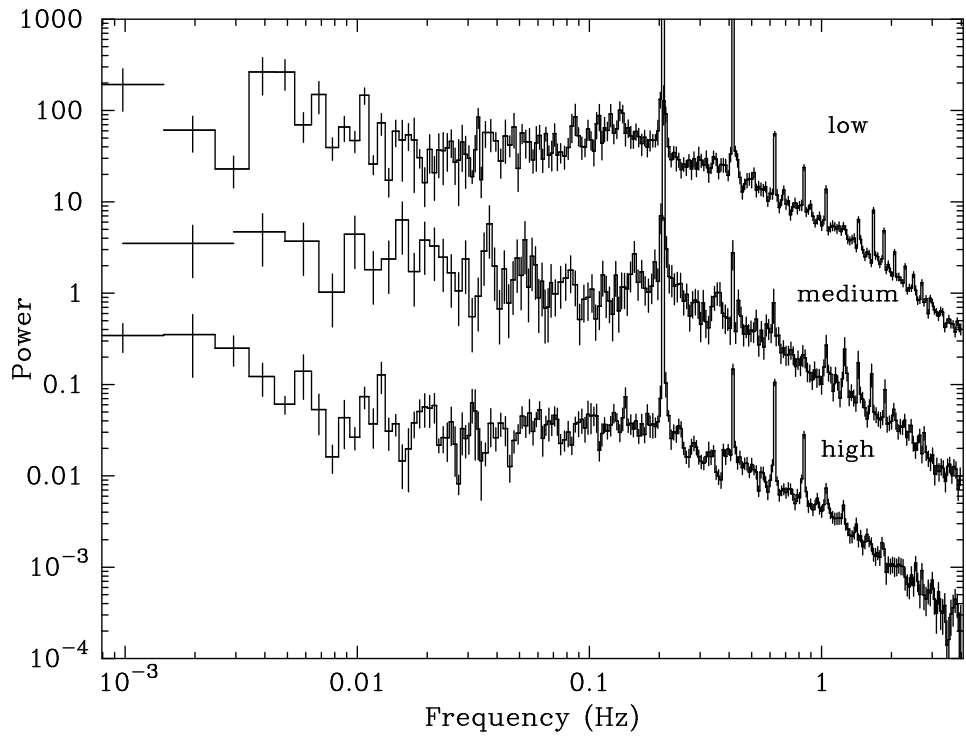}
\caption{Representative power spectra of Cen X-3 at different intensity state 
without QPOs are shown here. The first and second plots are multiplied by
constant numbers for clarity.}
\end{figure}

\begin{figure}
\centering
\includegraphics[height=5in, angle=-90]{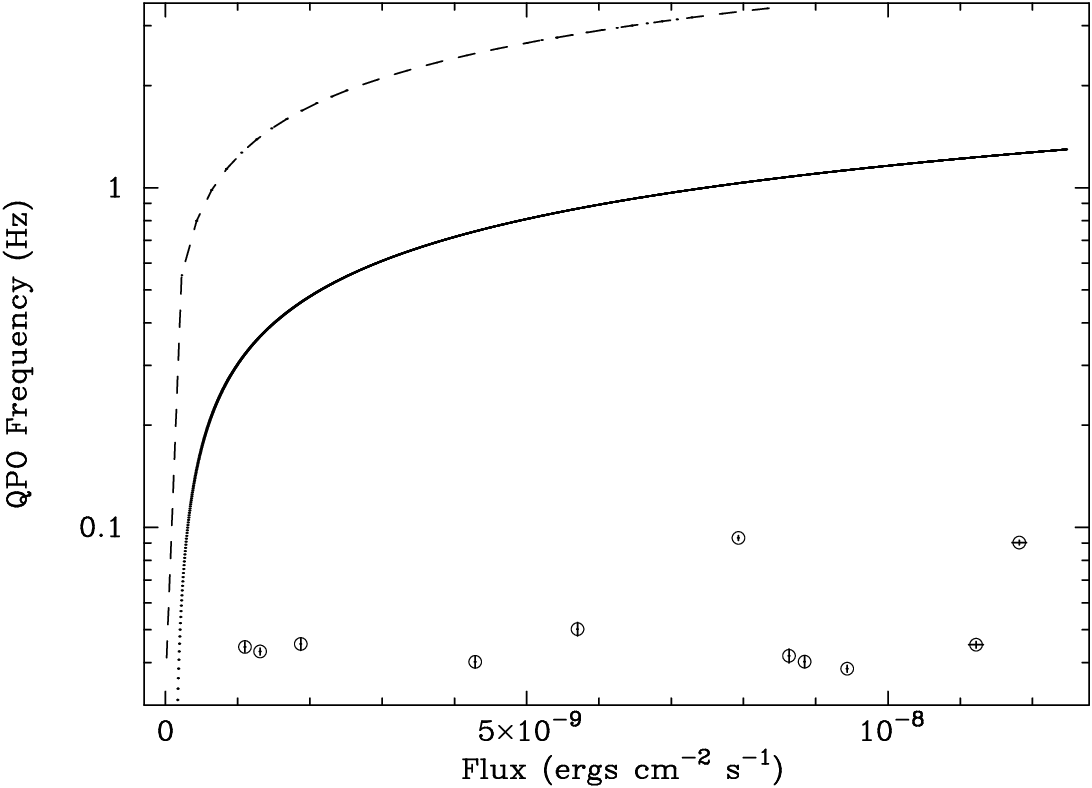}
\vskip 1cm
\includegraphics[height=5in, angle=-90]{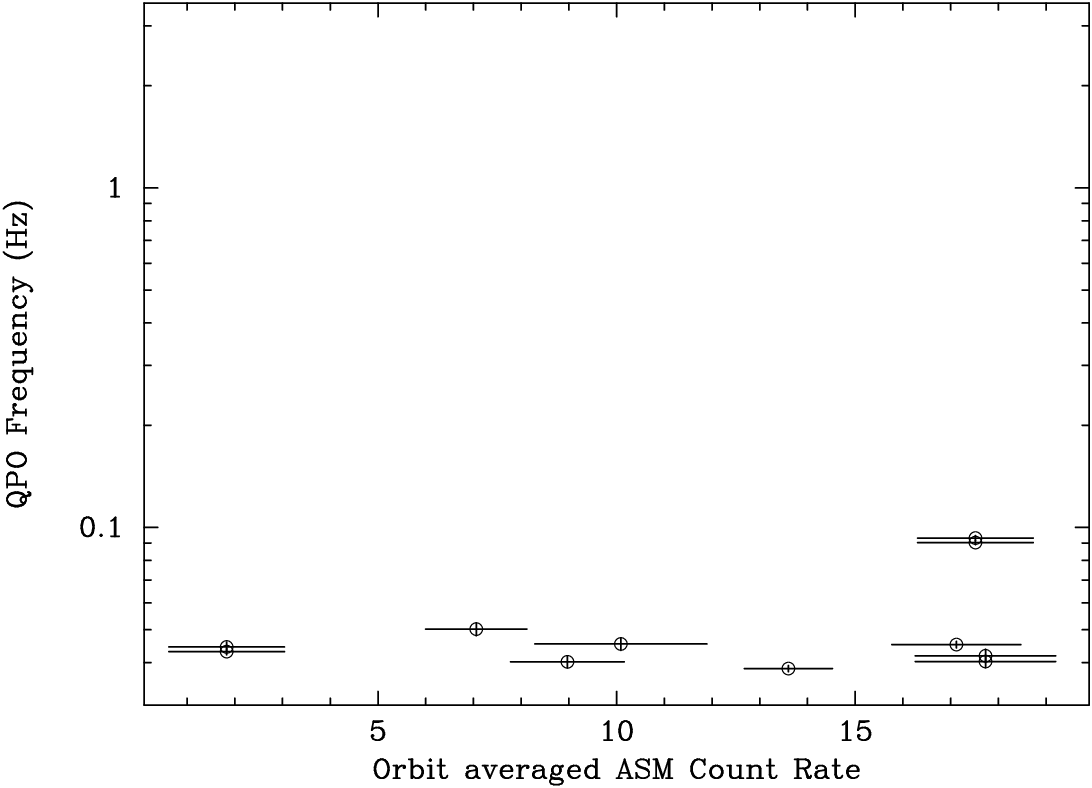}
\caption{Upper pannel plots (points marked with circles) the observed QPO 
frequency against the instantaneous flux as measured using the PCA spectrum. 
The dashed line shows the relation of QPO frequency and flux of the source as
expected due to the BFM when $r_M$ is calculated using the scaling factor of 
0.5 and solid line shows the same relation when $r_M$ is calculated without 
using the scaling factor. Lower panel plots the QPO frequency against the orbit
averaged ASM count rate of Cen X-3 at the time of observation.}
\end{figure}

\begin{figure}
\centering
\includegraphics[height=5in, angle=-90]{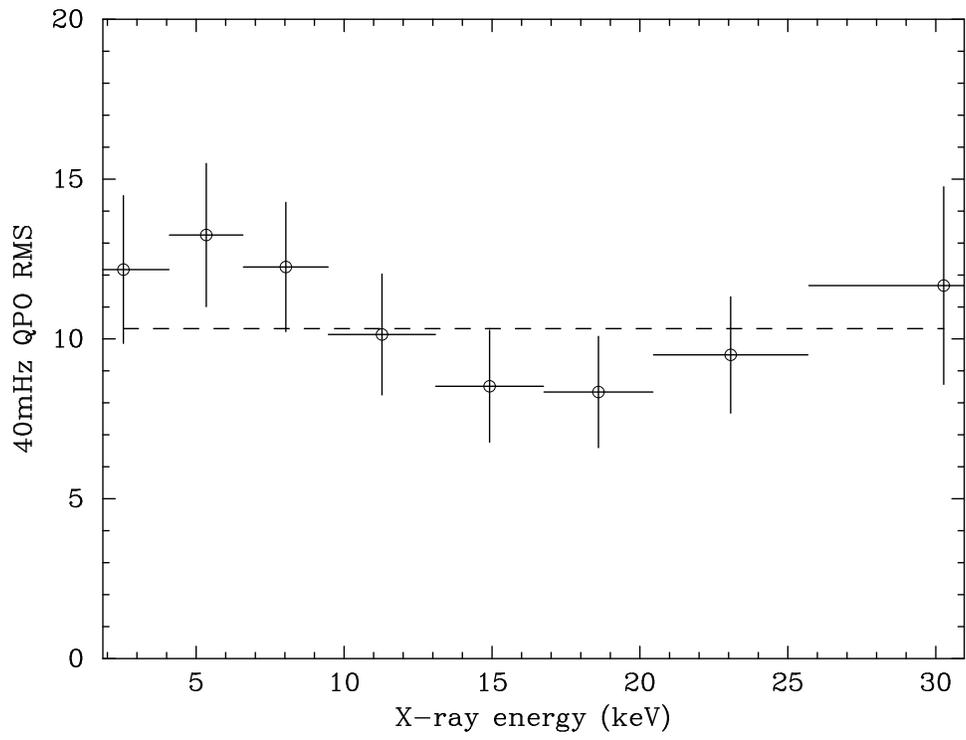}
\caption{RMS fluctuation in the 40 mHz QPO feature is shown here as a function
of energy for observation ID P10132. The average RMS value is shown with
the dashed line.}
\end{figure}

\end{document}